\documentclass[11pt]{article}
\usepackage{authblk}
\usepackage{orcidlink}
\usepackage[T1]{fontenc}
\usepackage{amsmath,amssymb}
\usepackage{graphicx}
\usepackage{lscape}
\usepackage{epsfig}
\usepackage[export]{adjustbox}
\usepackage[font=scriptsize]{caption} 
\usepackage{tikz}
\usepackage[subpreambles=true]{standalone}

\usepackage{hyperref}
\usepackage{color}

\usepackage{booktabs}

\title{Portrait comparison of binary and weighted\\Skill Relatedness Networks}
\author[1]{Sergio Andrés De Raco \orcidlink{0000-0002-1240-5799}
 \thanks{\href{mailto:sergioderaco@gmail.com}{sergioderaco@gmail.com}}}
\author[2]{Viktoriya Semeshenko \orcidlink{0000-0003-0295-5946}
 \thanks{\href{mailto:vika.semeshenko@gmail.com}{vika.semeshenko@gmail.com}}}
\affil[1]{Universidad de Buenos Aires. Facultad de Ciencias Económicas.
Buenos Aires, Argentina. Universidad de Buenos Aires. 
Instituto Interdisciplinario de Economía Política  de Buenos Aires. 
Buenos Aires, Argentina}
\affil[2]{Universidad de Buenos Aires. Facultad de Ciencias Económicas.
Buenos Aires, Argentina. 
CONICET-Universidad de Buenos Aires. 
Instituto Interdisciplinario de Economía Política de Buenos Aires. 
Buenos Aires, Argentina}
\date{}                     
\begin{document}
\maketitle 

\begin{abstract} 
In this paper we compare Skill-Relatedness Networks (SRNs) for selected countries, that is to say statistically significant inter-industrial interactions representing latent skills exchanges derived from observed labor flows, a kind of industry spaces. Using data from Argentina (ARG), Germany (DEU) and Sweden (SWE), we compare their SRNs utilizing an information-theoretic method that permits to compare networks of "non-aligned" nodes, which is the case of interest. For each SRN we extract its portrait, a fingerprint of structural measures of the distributions of their shortest paths, and calculate their pairwise divergences. This allows us also to contrast differences in structural (binary) connectivity with differences in the information provided by the (weighted) skill relatedness indicator (SR). We find that, in the case of ARG, structural connectivity is very different from their counterpart in DEU and SWE, but through the glass of SR the distances analyzed are all substantially smaller and more alike. These results qualify the role of the SR indicator as revealing some hidden dimension different from connectivity alone, providing empirical support to the suggestion that industry spaces may differ across countries. 
\end{abstract}


\section{Introduction}
\label{sec:intro}

Labor flows are a key factor in understanding economic activity, as they represent the interplay of workers' supply and firms' employment demand in the labor market. 
Particularly, job-to-job transitions are relevant labor flows, with recognized pro-cyclical behavior~\cite{mukoyama14} that carry tacit information about the relevance of past jobs' experience for new employers, specially those occurring between firms with different economic activities. These transitions are crucial for understanding the exchange of skills and abilities across sectors.

Traditionally, economists analyze labor flows with data at high level of aggregation of the standard classifications of productive activities, in order to correlate it with conventional national accounts data of sectoral activity.
The evolution of labor flows in Argentina has been analyzed using administrative records, which have shown that more disaggregated data can provide a richer picture of the temporal evolution of labor flows than aggregated data \cite{semeshenko_analysis_2021}. 
This is because employment flows carry information about the productive structure and diffuse knowledge among economic activities. Clearly, a more disaggregated level of detail, at the same time brings more complexity in interpretation tasks. 

Labor mobility across different industries reflects interconnections between economic activities, which can be effectively represented as networks. These networks highlight the properties of connectivity between economic sectors, offering insights into the flow of labor and the relationships between various industries within an economy. 

In Argentina, the Ministry of Labor, Employment and Social Security has data of administrative records of formal private labor employment from the Argentine Pension System\footnote{Spanish: Sistema Integrado Previsional Argentino (SIPA).} provided by the Observatory of Business and Employment Dynamics\footnote{Spanish: Observatorio de Empleo y Dinámica Empresarial (OEDE).}. 
The data includes interannual exchanges of employment between productive economic activities registered between 2009 and 2014. The set of activities includes nearly 400 sectors (branches of economic activities) at four digits of ISIC Rev.4 classifier. 

Previously, in ~\cite{de_raco_labor_2019, rdw19} the inter-industry labor flows of Argentina have been studied at high level of details, and revealed that networks extracted are typically very dense, not sparse, with clear core-periphery structures, and present small-world properties. Although these microscale networks provide new and useful information, they also pose several challenges for their interpretation and applications in, for example, policy design and analysis. The structure of interannual of labor Networks vary over time due to both cyclical and structural factors (\cite{de_raco_labor_2019}, \cite{semeshenko_analysis_2021}, \cite{deraco_semeshenko_2023}).
We also applied the skill-relatedness (SR) indicator measure for the analysis of labor flow dynamics \cite{neffke2017inter}, and compare it with the original flows in order to differentiate the type of information that each of these techniques offers for characterizing the productive system based on the dynamics of private formal employment \cite{de_raco_network_2019}. 

In this paper, we focus on the skill-relatedness networks (SRN). We are particularly interested in uncovering the structure of skill overlap between industries, as measured by labor flow transitions. To achieve this, we construct a network of normalized inter-industry labor flows following the methodology outlined in (Neffke et al. \cite{neffke2013skill}, and Straulino et al. \cite{straulino2021bi}) and characterize its structure.

Furthermore, we aim to compare the SRN of Argentina (ARG) with the SRNs of Germany (DEU) and Sweden (SWE). 
Our objective is to investigate to what extent the inter-industry labour networks differ between developing and developed countries?  
This comparison will provide insights into the differences in skill-relatedness patterns and industrial interactions across different economic contexts.

The proposed challenge translates into a new problem, because the underlying networks present systems of different dimensions, i.e. networks with non-aligned nodes. 
Comparing and identifying similarities between networks can indeed be a challenging problem. When given two networks, determining how similar they are typically involves quantifying their structural, topological, or functional similarities. Several methods and metrics have been developed to address this problem: Graph Invariants, Network Measures, Graph Matching Algorithms, Information-Theoretic Methods, Network Alignment, Machine Learning Approaches. Choosing an appropriate method depends on the specific characteristics of the networks and the research question at hand. 
Indeed, approaches to network comparison can be roughly divided into two groups based on whether they consider or require two graphs defined on the same set of nodes. When we consider networks defined on the same set of nodes, the comparison becomes straightforward since there's no need to align nodes between the two networks. For example, the cases of comparison of SRNs with the same number of nodes -aligned- has been already done by  \cite{straulino2021bi}. 
However, even if two networks have identical topologies, they might have no nodes or edges in common simply because they are defined on different sets of nodes. This highlights the importance of carefully considering the context and objectives when choosing a comparison approach for networks.

In the present case, we are dealing with a ``non-aligned`` network comparison, i.e. no nodes are necessarily shared between the networks. For this, we are using portraits divergence, a method for characterizing large complex networks by introducing a new matrix structure, unique for a given network, which encodes structural information, provides useful visualization, and allows for rigorous statistical comparison between networks \cite{bagrow2008portraits}.

The paper is organized as follows: in section \ref{sec:data} we described the tree datasets used in the analysis and methodology used, section \ref{sec:results} presents our results, and in section \ref{sec:conclu} we discuss results and research ahead. 
An ending appendix contains specific tables and graphs.

\section{Data and Methods}
\label{sec:data}

We utilize different datasets for selected countries: Argentina (ARG), Germany (DEU), and Sweden (SWE), at the level of 4 digits of detail of their national economic activity classifications, procured from various sources. 

These labor flows and skill-relatedness data are utilized to construct networks of interactions, while employment data is employed to determine the relative size of sectors. 
We call the interactions ``links'' or ``edges'' and the industries ``nodes''. 
Subsequently, we compare the built skill-relatedness networks using information-theoretic methods, specifically portraits divergence. 
A difficult problem when studying networks is that of comparison and identification, in particular, when they are defined on different sets of nodes, i.e. the size of the network is different, 
and thus, the number and/or economic activities of the underlying economic systems to be compared are different. 
These methods enable us to uncover structural information and conduct rigorous statistical comparisons between the networks.

\subsection{Data}

In order to build the Skill-Relatedness networks  (SRN)s for  ARG, SWE, and DEU we processed data at four digits of their national economic activity classifications (compatibles to \href{https://unstats.un.org/unsd/classifications/Econ/ISIC}{ISIC 4} or NACE 2 classification, the European version of ISIC 4) from different sources, explained herein and summarised in Table \ref{tab:table01}). 

In the case of Argentina, we utilize labor flow data for the period 2009-2014 obtained from the Observatory of Employment and Business Dynamics within the Ministry of Labor, Employment, and Social Security. This data is sourced from administrative records of the Federal Public Revenue Administration. 
With access to flow transition matrices, we proceed to calculate the skill-relatedness ($SR$) indicator, as outlined in \cite{neffke2013skill}, following the methodology described in the next section. Subsequently, we construct the corresponding Skill-Relatedness Networks (SRNs).
 
In the case of Germany, we use directly the $SR$ data at four digit WZ08 national industrial classification (equivalent to NACE 2), for the period 2007-2013, published in \cite{neffke2013skill} by the authors\footnote{See ``Skill relatedness matrices for Germany'' at \url{https://iab.de/publikationen/publikation/?id=7202046}.} originally estimated from data of the Employee History\footnote{German: Beschäftigten-Historik, BeH.}, based on the social security records of Germany. 
Additionally, we use German employment data from \href{https://www.destatis.de/}{DESTATIS}, the Federal Statistical Office of Germany.  

In the case of Sweden, we use directly the $SR$ data at four digit SNI 2007 national industrial classification (equivalent to NACE 2), for the period 2007-2017, calculated by the Swedish Agency for Growth Policy Analysis (\cite{sweden2021}) using the methods in \cite{neffke2013skill} with Swedish administrative data\footnote{See ``Skill relatedness matrices for Sweden'' at \url{https://www.tillvaxtanalys.se/in-english/publications/pm/pm/2021-05-18-skill-relatedness-matrices-for-sweden.html}.}. 
We use Swedish employment data from \href{https://www.scb.se/en_/}{Statistics of Sweden} for the period of analysis.

\begin{table}
\small
    \centering
    \begin{adjustbox}{width=\textwidth}
    \begin{tabular}{|l|p{4cm}|p{4cm}|p{4cm}|}
         \hline
             & \textbf{Argentina} & \textbf{Germany} & \textbf{Sweden}\\
         \hline
         \hline
        \textbf{Data} & Inter-industry labor flows & Inter-industry skill-relatedness & Inter-industry skill-relatedness\\
         \hline
        \textbf{Classification} & ISIC 4 & WZ08 (NACE 2) & SNI 2007 (NACE 2)\\
         \hline
        \textbf{Period} & 2009-2014 & 2007-2014 & 2007–2017\\
        \textbf{Years (\#)} & 5 & 7 & 10\\
         \hline
        \textbf{Flows} &  &  & \\
        . total & 2,060,515 & 5,529,890 & 5,100,000\\
        . avg./year & 412,103 & 789,984 & 510,000\\
         \hline
        \textbf{Sectors (\#)} &  &  & \\
        . original & 410 & 597 & 586\\
        . SR+ & 407 & 584 & 577\\
         \hline        
        \textbf{Source} & Ministry of Labor, Employment, and Social Security, based on Federal Public Revenue Administration data & \cite{neffke2017skill}, Table 2, based on Beschäftigten-Historik, Federal Statistical Office & Rapport 2021:02:04, Swedish Agency for Growth Policy Analysis, based on LISA data, Statistics of Sweden\\
         \hline
        \textbf{URL} & N/A & DESTATIS.de & SCB.ce \\
         \hline
        \end{tabular}
         \end{adjustbox}
    \caption{Data reference summary for Argentina, Germany and Sweden. Administrative data at 4 digits of economic activity classifications. Comparable classification systems ISIC 4 and NACE 2. In relative terms with respect to total and average employment, the differences in inter-industry flows in Germany and Sweden with respect to  Argentina are due to the larger number of nodes in the networks, and larger periods.}
    \label{tab:table01}
\end{table}

\subsection{Methods} 

\begin{figure}[ht]
    \centering
    \includegraphics[width=1\linewidth]{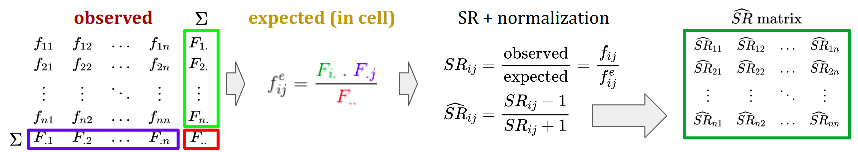}
    \caption{The scheme of construction of skill-relatedness indicator used for Argentina. The sequential steps from the observed flow matrix to the skill relatedness indicator matrix. A reference matrix of ``expected flows'', $f^{e}_{ij}$, built on the basis of the edges (eg: totals per rows, $F_{i.}$, columns, $F_{.j}$, and table, $F_{..}$) of the matrix of observed flows. This matrix reflects ``random'' flows in the sense that sectoral exchanges are proportional to the outflows and inflows between sectors with respect to total flows. For each cell an associated matrix of elements, $SR_{ij}$, is calculated as the ratio of the observed value of employment flows with respect to the theoretical or expected value. Thus, one can interpret values less than unity, $SR_{ij}\in[0,1)$ as not moving away  from a random distribution significantly, while values greater than unity, $SR_{ij}\in[1, + \infty)$, showing deviations from the proposed random distribution as benchmark. The SR matrix is symmetrized by means of averaging the SR matrix with its transpose. In this way the related graph becomes undirected.}
    \label{fig:SR-indicator}
\end{figure}

\noindent\textbf{Skill-Relatedness Networks.}
Given that on the Argentine side we have the data of flows, i.e. transition flows matrices, we first proceed to construct the skill-relatedness networks according to the methodology outlined in  \cite{straulino2021bi}. 
We calculate the skill-relatedness indicator, $SR_{ij}, \forall i,j \in N$, where $N$ represents the total number of industries (hereafter used interchangeably with "economic activities" or "sectors") included. 
The skill-relatedness indicator between industries $i$ and $j$ is computed as a ratio between the \textit{observed} labor flows and the \textit{expected} flows from a null model, which is calculated from the margins of the respective  ($A_{N\times N}$) flow matrix for each cell (see \cite{neffke2013skill}, \cite{neffke2017inter}, and \cite{neffke2017skill} for further insights into this methodology), see Fig. \ref{fig:SR-indicator}. The indicator is then symmetrized and normalized to map it to the interval $SR\in[-1,1]$.

\begin{figure}[!ht]
    \centering
    \includegraphics[width=1\linewidth]{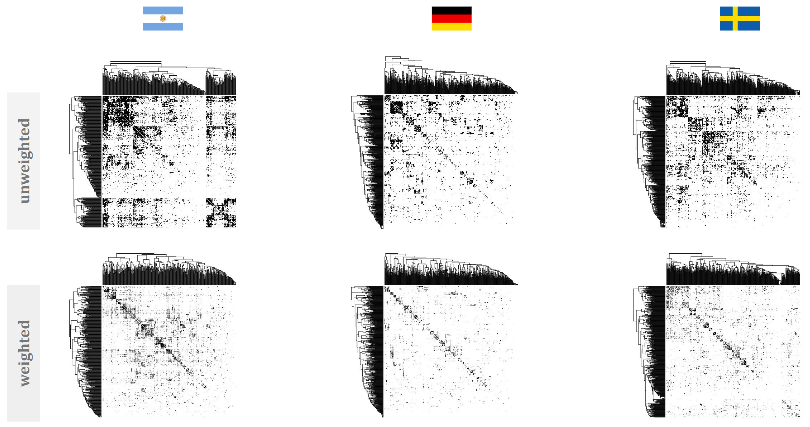}
    \caption{Skill-Relatedness Networks (SRNs). Heatmap representation of undirected filtered (positive) networks: Unweighted (binary, upper row) and weighted (lower row) SRNs. Sorting is done with a hierarchical clustering algorithm with complete linkage. Visualizations of $SRN^{+}$ for Argentina (ARG), Germany (DEU), and Sweden (SWE) for periods and size according to the specifications in Table \ref{tab:table01}. }
    \label{fig:sr-networks}
\end{figure}

In the cases of Germany and Sweden, we count with skill-relatedness data to build the matrices directly. For further analysis and network comparison,  we keep only positive values of skill-relatedness values of the matrices.  
Bounding to positive values, seems to be an appropriate method and a proper criteria for pruning the networks of the less significant flows in the ``skill-relatedness'' sense. 
Values greater than 0 indicate that the number of observed job switches is greater than what would be expected at random under the null model specified, i.e. workers that would have moved at random given the respective size of each industry (similar to the \emph{Configuration Model}).
Hereafter we refer to these networks with positive skill-relatedness, $SR_{ij}>0$, as $SRN^{+}$s or simply $SRN$s and conveniently index them by country whenever needed \cite{straulino2021bi}. 
Fig  \ref{fig:sr-networks} shows the skill-relatedness networks for three respective datasets. We plot the heatmap respresentation for both, unweighted, ie. binary network (Fig.  \ref{fig:sr-networks}, lower row), and weighed networks (Fig.  \ref{fig:sr-networks}, upper row). 

Regarding the size of the networks to compare, which refers to the number of nodes or industries included in the analysis, it's worth noting that Germany and Sweden have more than 40\% more industries at their four-digit detailed classification compared to Argentina.
This difference in network size presents the challenge of comparing networks that are ``non-aligned'', meaning they have different numbers of nodes, where no nodes are necessarily shared between the networks. 
To address this issue, we employ the information-theoretic method of portrait network divergence, which was developed in \cite{bagrow2019information}, and  is suitable for comparing networks of different sizes and without node  correspondence. 
A common approach for comparison without assuming node correspondence is to utilize a comparison measure based on a \textit{graph invariant}. Graph invariants are properties of a graph that hold for all isomorphisms of the graph. 
Using an invariant helps alleviate concerns about the encoding or structural representation of the graphs, allowing the corresponding measure to focus solely on the topology of the network. Graph invariants can take various forms, including probability distributions.
Thus, by focusing on the topology of the networks and abstracting from the problem of node correspondence, we can compare these networks, without ensuring that networks use the exact same industrial classification encoding, which allows for a direct comparison of their structures without the need to align nodes. This approach enables us to analyze the similarities and differences in the network topology across different countries or contexts.

\begin{figure}[ht]
    \centering
    \includegraphics[width=1\linewidth]{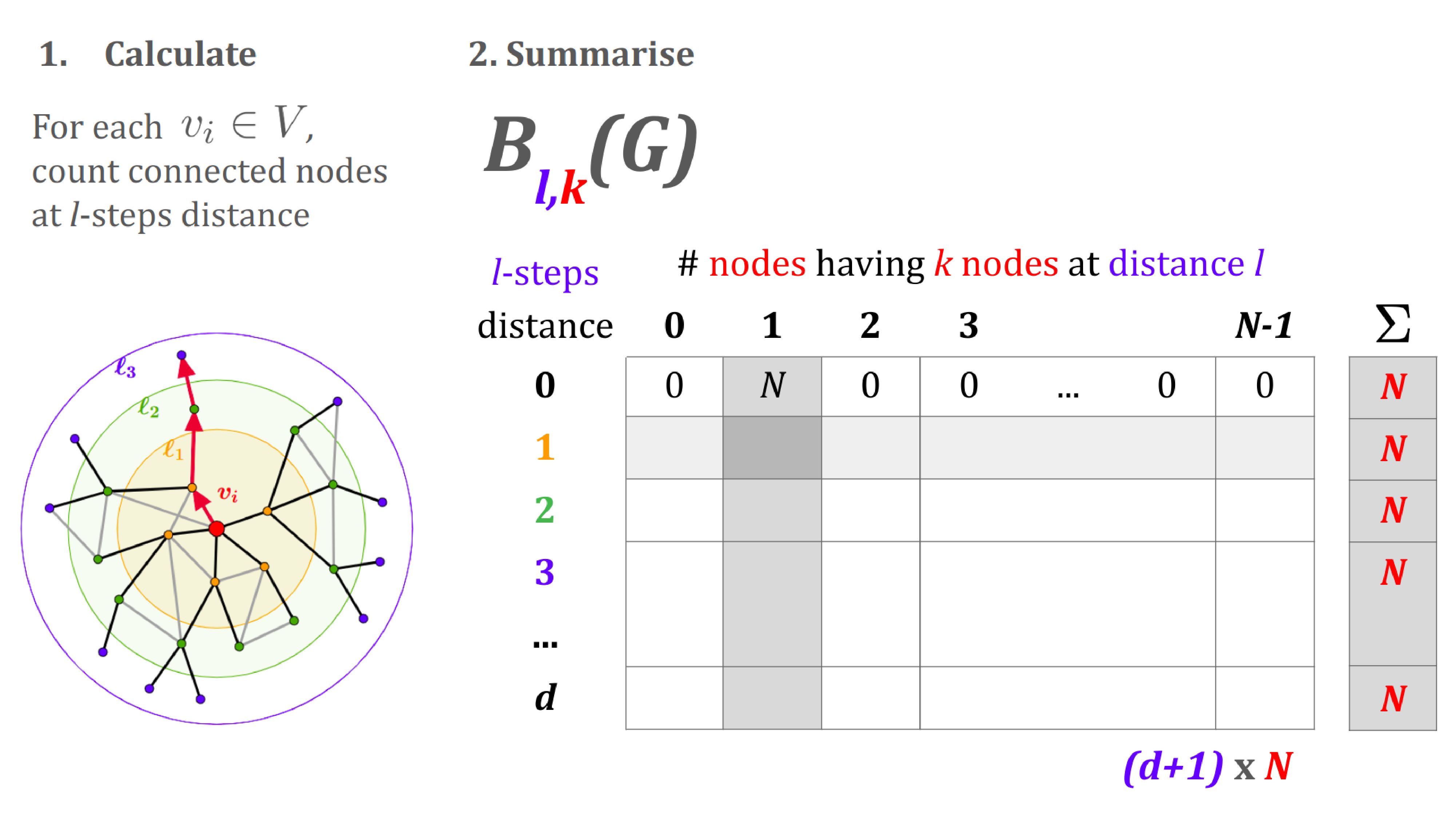}
    \caption{$B_{\ell,k}(G)$-matrix construction. For each node $v_i\in V$, count connected nodes at $\ell$-steps distance, its $\ell$-shell or connectivity layer, then summarise for each $\ell$-distance (in rows) the number of nodes that have $k$-neighbors, taken as shortest path length ($\ell$-shells). The first row $\ell=0$ gives the number of nodes. The second row $\ell=1$ stands for degree distribution: each sector’s number of direct connections. The subsequent rows $\ell \geq 2$ distribution of l-nearest neighbours. The last row $\ell=d$ gives the diameter of the network, i.e. longest shortest path in the network. }
    \label{fig:b-mat}
\end{figure}

\noindent\textbf{Portraits.}
The method stands on the construction of a $B_{\ell,k}$-matrix (v.g.: the network portrait, see \cite{bagrow2008portraits}) consisting of:
$$B_{\ell,k}\equiv\text{ the number of nodes who have (exactly) }k\text{ nodes at distance }\ell$$
for $0\leq\ell\leq d$ and $0\leq k\leq N-1$, where the distance is taken as the shortest path length and $d$ is the graph's diameter (see Fig.~\ref{fig:b-mat}). 
In this sense, like onion layers, each node $v_i$ is surrounded by $\ell$-shells or connectivity layers of order $\ell$.
The rows represent histograms (or distributions) of $\ell$-order shortest paths. This matrix condenses structural properties of the network based on the distance connecting two nodes in terms of successive links or path lengths, $\ell$, which encode shortest path distributions, for example including the degree distribution ($\ell=1$, for an unweighted network) and higher order paths. It is important to state that the network portraits are agnostic of the identity of the nodes, capturing topological information without reference to the nodes attributes. As a graph invariant, the $B$-matrix of a network is unique and can be used as a network ``fingerprint''. In this way, comparing two networks $G$ and $G'$ can be translated into comparing their portraits, $B$ and $B'$. 

\noindent\textbf{Network Portrait Divergence.}
After computing the portraits of these networks, say $G$ and $G'$, each portrait can be transformed into matrices of row-wise probability distributions, then reduce them to two single joint distributions for all rows which can be used to define a single Kullback-Liebler (KL) divergence between their portraits (see \cite{bagrow2019information}). The  network portrait divergence (NPD) is defined then as the Jensen-Shannon divergence:
$$ D_{JS}(G,G')\equiv\frac{1}{2}KL(P||M)+\frac{1}{2}KL(Q||M), \in [0,1]$$
where $M\equiv\frac{1}{2}(P||Q)$ is the mixture distribution of $P$ and $Q$, where $P$ is $P(k,\ell)=\frac{kB_{\ell,k}}{N^2}$ and $Q$ is, likewise, $Q(k,\ell)=\frac{kB'_{\ell,k}}{N^2}$. 

Note that unweighted networks will have integer diameter $d$, while for weighted networks $d\in\mathbb{R}$ is continuous. In this latter case, which specifically concern us for the comparison of SRNs, the shortest paths may have non-integer paths so the algorithm used to find the shortest paths for unweighted networks changes from breadth-first-search to Dijkstra's algorithm. Also a binning strategy for aggregating (continuous) shortest paths is due. A simple one is to use $b$ bins as quantiles to be able to compute the portraits, $B_{\ell,k}$ and $B'_{\ell,k}$, of each network. In our case we choose binedges regarding the weight distribution of the SRNs under analysis.

\section{Results}
\label{sec:results}

The SRNs for each country, built from the positive skill-relatedness indicator matrices and ancillary employment data, present a visible dense structure with a unique giant component (induced by the construction of $SRN^{+}$) having short paths and diameter (top row in Fig.~\ref{fig:sr-networks}). 
As reported in Table~\ref{tab:table01}, although the statistical systems of classification for economic sectors where compatible between countries,
the size of these networks vary because of: a) differences in some sectors' specification as informed by each country, and; b) as a result of the filtering process described in section \ref{sec:data}, of significantly observed flows in terms of the skill-relatedness criteria.

\begin{figure}[ht]
    \centering
    \includegraphics[width=1\linewidth]{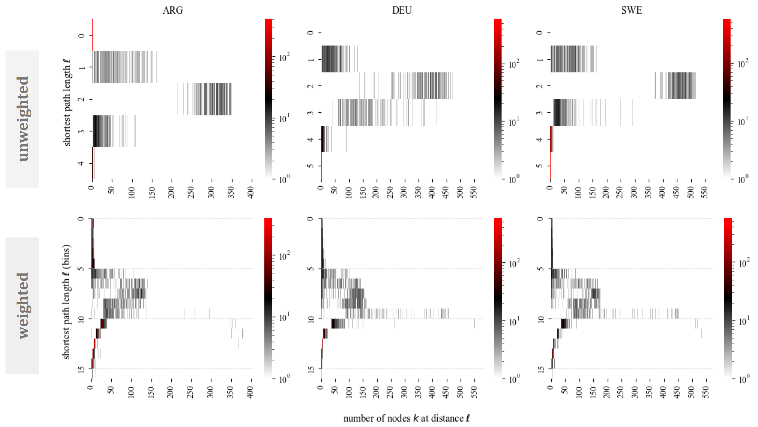}
    \caption{Network portraits. Upper row: Unweighted (binary) SRNs. Discrete shortest path length $\ell$ from 0 to $d$, the diameter of the network with $d_{ARG}=4$, and $d_{DEU}=d_{SWE}=5$.
    Lower row: Weighted SRNs. Continuous (binned) shortest path length $\ell$ from 0 to $d$, the diameter of the network with $d_{ARG}=1.95$, $d_{DEU}=1.43$, and  $d_{SWE}=1.96$. For a better visualization we used 16 bins (vertical axis), with smaller bins destined to lower values of $SR>0$ 
    (binedges $\in (0, 0.002, 0.004, 0.006, 0.008, 0.01, 0.02, 0.03, 0.04, 0.05, 0.1, 0.25, 0.5, 0.75, 1, 1.5,2)$). 
    }
    \label{fig:B-matrices}
\end{figure}

After building each country SRN, we computed their respective portraits for weighted, $Bw_{\ell,k}^{c}$, as well as unweighted, $B_{\ell,k}^{c}$, versions of the SRNs with $c\in\{ARG, DEU, SWE\}$, plotted in Fig. \ref{fig:B-matrices}. We use their unweighted versions to naturally introduce a way to better comprehend the information contained therein in terms of node connectivity. 

In a network portrait, $\ell$ refers to the length of shortest paths and $k$ counts the ``number of nodes`` having paths of length $\ell$, that is to say considering $\ell$-shells of each node in the network (see Fig.\ref{fig:b-mat}). 
In an unweighted network $\ell=1$ is the degree distribution, $\ell=n$ is the distribution of shortest paths of order $n$, and $\ell=d$ is the max length representing the network diameter. In a weighted network, $\ell$ has to be discretised as it is continuous. 

For the \textit{unweighted portraits} (upper row in Fig. \ref{fig:B-matrices}), depicting the fingerprints of the pure connectivity in the SRNs, show the distribution of shortest paths for each country's network. 
These portraits present a kind of $P$-shape related to the big connected component topology that is characteristic of SRNs, as mentioned earlier.
Their range goes from $\ell=0$ (representing the total count number of nodes, $N$), occurring $B_{\ell,k}^{c}=N_c$ for each country network, to $\ell=d$, the corresponding (unweighted)  diameter of each network (v.g.: $d_{ARG}=4$, and $d_{DEU}=d_{SWE}=5$). Intuitively, the visualizations of this portraits show a condensed image of the way nodes, economic sectors in SRNs, are connected and proximate to each other albeit not identifying the specific connection between any pair of sectors $m$ and $j$. The second row ($B_{\ell=1,k}^{c}$) corresponds naturally to the standard degree distribution of direct connections. It can be appreciated that this distribution is relatively more widespread for ARG than for SWE and DEU, with DEU accumulating relatively more (less connected) nodes in small values of $k$, that is to say more nodes with small $\ell$ order direct neighbourhoods. The next row, $B_{\ell=2,k}^{c}$, show the distribution of ``two steps'' paths or the most proximate indirect neighbourhood shell for each node ($\ell$-shell=2), that is to say: industries connected (through $SR$-links) with the industries in their direct connections circle. It can be appreciated that all countries show distributions centred in higher values of $k$, corresponding to the majority of nodes (industries) having a great number of nodes (industries) at this distance. In this case, DEU has a relatively more widespread distribution, while ARG and SWE appear more alike with higher density in high values of $k$. This means that most sectors show many ``two steps'' connections, a fact consistent with the analysis of labour flow networks for Argentina evidencing dense networks with short average paths and diameter, and having small world properties (v.g.: typical diameter of three steps, see \cite{de_raco_labor_2019}, \cite{de_raco_network_2019}). The following row, $B_{\ell=3,k}^{c}$, show the distribution of ``three steps'' paths length, an enhanced indirect neighbors set. It can be appreciated that the distributions are again skewed towards lower values of $k$, meaning that as the length of shortest paths approaches the diameter (shortest paths maximum length) there are less nodes (industries) having many nodes at this distance. In this case, ARG and SWE appear more similar with a greater concentration of nodes (industries) having a small $k$ number of nodes at a three step distance, while DEU has more dispersed distribution with higher values of $k$ nodes at three steps distance. This suggests that DEU has deeper chains of connectivity, say showing more cohesion, than ARG and SWE. The last rows of these unweighted portraits, referring to the more distant layers of connectivity near or at their (respective) diameters, show high concentration of these longer paths in lower values of $k$. This refers to the paths linking nodes with sectors in the outer periphery having very poor connectivity. 

\begin{figure}[ht]
    \centering
    \includegraphics[width=0.85\linewidth]{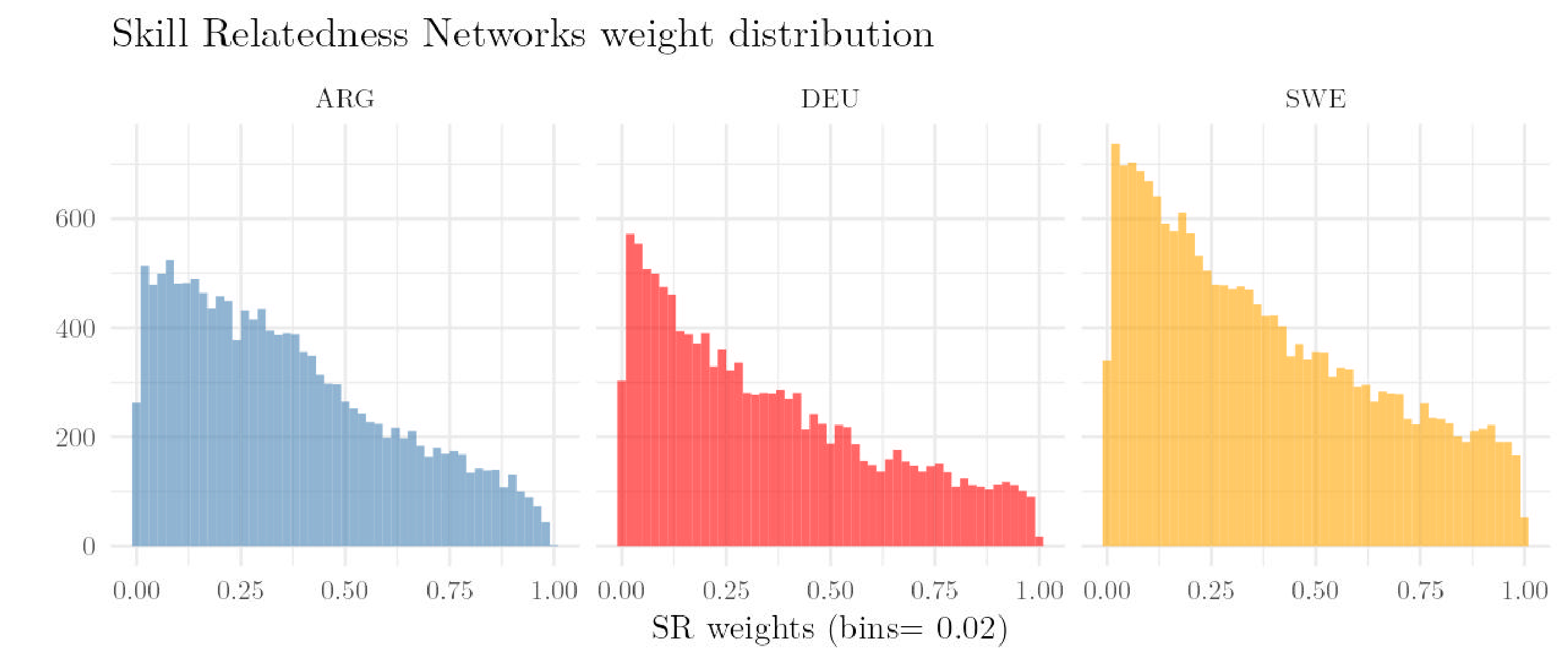}
    \caption{SRNs weight distribution.}
    \label{fig:weights}
\end{figure}

For the \textit{weighted portraits} (lower row in Fig. \ref{fig:B-matrices}), depicting the valued fingerprints of the SRNs, show the distribution of shortest paths in terms of $SR$ for each country's network. Their range goes from $\ell=0$  to $\ell=d$, in this case corresponding to the continuous diameter of each network (v.g.: $d_{ARG}=1.95$, $d_{DEU}=1.43$, and $d_{SWE}=1.96$). To compare this portraits showing the distributions of weighted shortest paths,  we computed the same number of bins for the three SRNs so the interpretation can equally be made for all values of (binned) $\ell$. 
As can be appreciated, the interpretation of weighted path lengths and the comparison between them is more demanding although differences and similarities can be appreciated between the fingerprints. The chosen binning, with binedges $\in (0, 0.002, 0.004, 0.006, 0.008, 0.01, 0.02, 0.03, 0.04, 0.05, 0.1, 0.25, 0.5, 0.75, 1, 1.5,2)$, highlights lower $SR+$ weights in line with their decreasing prevalence in SRNs (see Fig.~\ref{fig:weights}) across the (maximum) range, $r\in(0,max(d_c))$, of observed weighted paths for all countries.

With this weight aggregation, the weighted portraits in the lower row of Fig.~\ref{fig:B-matrices} can be divided into three ``charge zones'' in relation with the quantification of sector connectivity referenced in the horizontal axis and the weighted paths measured in the vertical axis: 
\begin{enumerate}
    \item [a] \textit{high-concentration low-weighted shortest} $\ell$-paths in \textbf{bins 1 to 5}, corresponding to a total weighted distance of $\ell \in (0.00,0.01)$ and involving the interconnection of just a few sectors; 
    \item [b] \textit{high-dispersion medium-weighted}  shortest $\ell$-paths in \textbf{bins 6 to 10}, corresponding to a total weighted distance of $\ell \in [0.01, 0.10)$, involving a sharply increasing interconnected (horizontal dispersion) and decreasing concentration (low intensity, showed in black and white gradient colors) sectors topology; and 
    \item [c] \textit{high-concentration high-weighted} shortest $\ell$-paths in \textbf{bins 11 to 16}, corresponding to a total weighted distance of $\ell \in [0.10,2.00]$ and involving the decreasing interconnection of a sectors
\end{enumerate}

\begin{figure}[ht]
    \centering
    \includegraphics[width=1\linewidth]{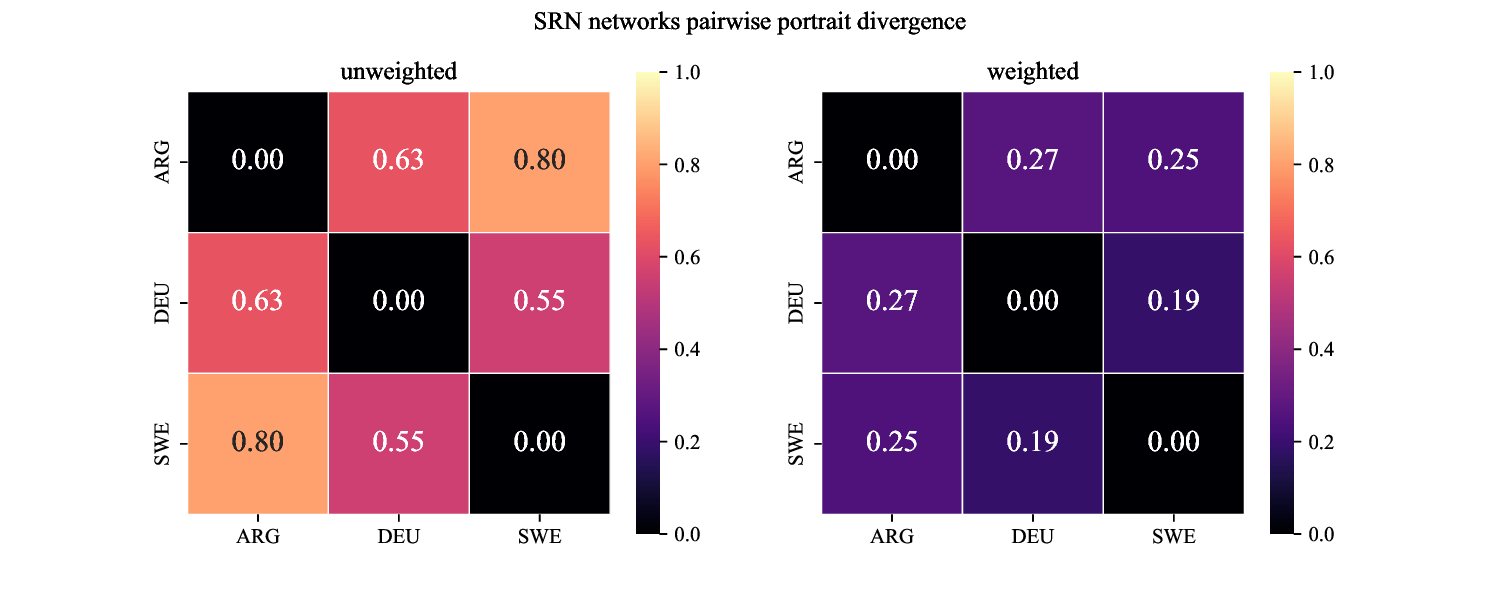}
    \caption{SRNs network pairwise portrait divergence. Right: weighted portraits divergence. Left: unweighted portraits divergence. Color range for $D_{JS}(G,G')\in[0,1]$, greater values showing more dissimilar network portraits.}
    \label{fig:npdivergence}
\end{figure}

To quantify these dissimilarities we calculate the (pairwise) network portrait divergence, $D_{JS}(G,G')\in[0,1]$, with higher values showing more dissimilarity, presented in Fig.~\ref{fig:npdivergence} for both unweighted and weighted portraits. 
The comparison for the \textit{unweighted portraits} present stark differences between ARG and those of DEU (0.63) and SWE (0.80), while at the same time it is also informative of the differences between DEU and SWE structure (0.55). In light of this results, it is useful to revisit the original binary structure of the SRNs in the upper row of Fig.~\ref{fig:sr-networks}. Taking the case of ARG, it is quite clear that its (clustering ordered) connectivity structure differs strikingly with both DEU and SWE. In particular, in ARG there is a group of approximately 30\% of total sectors (bottom right) with high interconnection within them and some non-trivial interconnection with the rest of the sectors. In turn the rest of the sectors are grouped and ordered in decreasing order of total connectivity, showing some subgroups with more connectivity within. In the case of DEU, the connectivity structure is smoothly decreasing and characterized by a small modular structure, with some small sector groupings with high connectivity within. The case of SWE appears as an intermediate between the others, also with smoothly decreasing modular connectivity structure but with subgroups bigger than in the case of DEU. 

Regarding the comparison of \textit{weighted portraits}, the differences of ARG's SRN and their counterparts in DEU (0.27) and SWE (0.25) appear less pronounced, and the comparison between DEU and SWE (0.19) show the lowest divergence. Again, it is useful to revisit the original weighted SRNs in the lower row of Fig.~\ref{fig:sr-networks}. This time the visible connectivity structure is more difficult to disentangle because of the weak density in all cases. In particular, SWE shows presents more modular structure detectable with the hierarchical clustering at the corners up-left (higher intersectoral connectivity within and also between this group and the immediate neighbors down/right, more central), and down-right (smaller group, less connected with the rest of the network, more periphery like).

%
\section{Discussion}
\label{sec:conclu}

In this paper we presented a comparison of different countries' skill-relatedness networks (SRNs) using data from Argentina (ARG), Germany (DEU) and Sweden (SWE) to assess the possible differences between SRNs in a developing country vis-a-vis those in developed countries. To this end we used a method suitable to compare networks of different size (non-aligned networks) that focuses on topological information \cite{bagrow2019information} using a measure of network portraits \cite{bagrow2008portraits}, a condensed representation of shortest path length structural information that compose a unique network fingerprint. Through this applied exercise we found that the methods of portrait representation of networks and the measure of network portraits divergence appear as appropriate methods to characterize and compare SRNs. 

We found that the comparison of \textit{unweighted} portraits of these networks show contrasting differences between the pure connectivity (binary) structure of the SRNs of a developing country like Argentina to the corresponding SRNs for developed countries like Germany and Sweden, with stark differences of interindustry connectivity in Sweden and high contrast with Germany's. Moreover, this comparison reveals important differences in the connectivity structure between Germany and Sweden networks. When comparing the more relevant \textit{weighted} skill-relatedness networks we found less contrasting differences between all the SRNs. In particular, Argentina's SRN appears quite dissimilar to the corresponding to both DEU and SWE, and at the same time Germany's SRN is quite similar to that of Sweden. These preliminary findings may give relative support to the hypothesis of similarity of different countries SRNs conditioned on historical and cultural differences (see \cite{neffke2013skill}). On the other hand, they show that the connectivity (topological) structure of different observed SRNs present stark differences between countries.


\bibliographystyle{unsrt}
\bibliography{asaid_arxiv_2024}

\begin{thebibliography}{10}

\bibitem{mukoyama14}
Toshihiko Mukoyama.
\newblock The cyclicality of job-to-job transitions and its implications for aggregate productivity.
\newblock {\em Journal of Economic Dynamics and Control}, 39:1--17, 2014.

\bibitem{semeshenko_analysis_2021}
Viktoriya Semeshenko and Sergio~A. De~Raco.
\newblock Analysis of the evolution of labor market flows in {Argentina}.
\newblock {\em Proceedings 50 JAIIO-AGRANDA}, pages 20--24, 2021.

\bibitem{de_raco_labor_2019}
Sergio~A. De~Raco and Viktoriya Semeshenko.
\newblock Labor mobility and industrial space in {Argentina}.
\newblock {\em Journal of Dynamics \& Games}, 6(2):107, 2019.

\bibitem{rdw19}
Sergio~A. De~Raco and Viktoriya Semeshenko.
\newblock The network structure of inter-industry labor mobility in {Argentina}.
\newblock In {\em 6th Regulating for Decent Work Conference}, Geneva, 2019. ILO.

\bibitem{deraco_semeshenko_2023}
A.~De~Raco, Sergio and Viktoriya Semeshenko.
\newblock Identificación de diferencias y similitudes estructurales de las redes interindustriales del empleo de argentina.
\newblock {\em to appear in Proceedings 52 JAIIO-AGRANDA}, 2024.

\bibitem{neffke2017inter}
Frank~MH Neffke, Anne Otto, and Antje Weyh.
\newblock Inter-industry labor flows.
\newblock {\em Journal of Economic Behavior \& Organization}, 142:275--292, 2017.

\bibitem{de_raco_network_2019}
Sergio~A. De~Raco and Viktoriya Semeshenko.
\newblock The network structure of inter-industry labor mobility in {Argentina}.
\newblock In {\em 6th Regulating for Decent Work Conference}, Geneva, 2019. ILO.

\bibitem{neffke2013skill}
Frank Neffke and Martin Henning.
\newblock Skill relatedness and firm diversification.
\newblock {\em Strategic Management Journal}, 34(3):297--316, 2013.

\bibitem{straulino2021bi}
Daniel Straulino, Mattie Landman, and Neave O’Clery.
\newblock A bi-directional approach to comparing the modular structure of networks.
\newblock {\em EPJ Data Science}, 10(1):13, 2021.

\bibitem{bagrow2008portraits}
James~P Bagrow, Erik~M Bollt, Joseph~D Skufca, and Daniel Ben-Avraham.
\newblock Portraits of complex networks.
\newblock {\em Europhysics letters}, 81(6):68004, 2008.

\bibitem{sweden2021}
SAGPA.
\newblock Skill relatedness matrices for sweden.
\newblock Technical report, Swedish Agency for Growth Policy Analysis, 2021.

\bibitem{neffke2017skill}
Frank Neffke, Anne Otto, Antje Weyh, et~al.
\newblock Skill-relatedness matrices for {Germany}: Data method and access.
\newblock Technical report, Institut f{\"u}r Arbeitsmarkt-und Berufsforschung (IAB), 2017.

\bibitem{bagrow2019information}
James~P Bagrow and Erik~M Bollt.
\newblock An information-theoretic, all-scales approach to comparing networks.
\newblock {\em Applied Network Science}, 4(1):1--15, 2019.

\end{thebibliography}

\appendix
\section{Appendix}\label{appendix}

Figures \ref{fig:ARG-srn}, \ref{fig:DEU-srn}, and \ref{fig:SWE-srn} provide detailed visualizations of SRNs for each country.

\begin{landscape} 
\begin{figure}[ht]
    \caption{\footnotesize{Argentina Skill Relatedness Network 2009-2014. Colors represent ISIC rev. 4 economic industrial classification, numbers in color legend are number of sectors under the correspondent letter division. Node size represents average sectoral employment. Edge width represents $SR>0$.}}
    \centering
    \includegraphics[width=\pdfpagewidth]{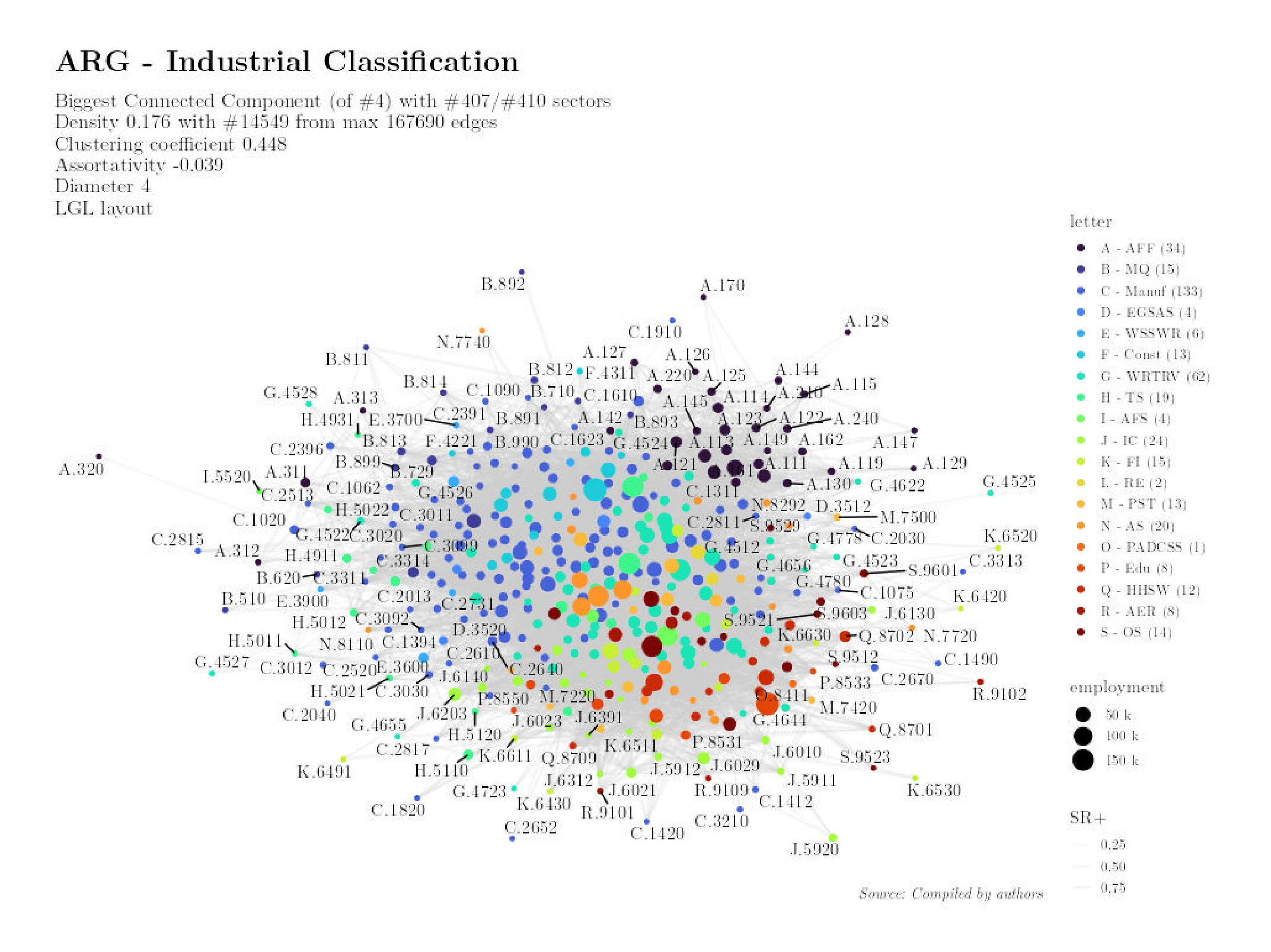}
    \label{fig:ARG-srn}
\end{figure}

\begin{figure}[ht]
    \caption{\footnotesize{Germany Skill Relatedness Network 2007-2014. Colors represent ISIC rev. 4 equivalent economic industrial classification, numbers in color legend are number of sectors under the correspondent letter division. Node size represents average sectoral employment. Edge width represents $SR>0$.}}
    \centering
    \includegraphics[width=\pdfpagewidth]{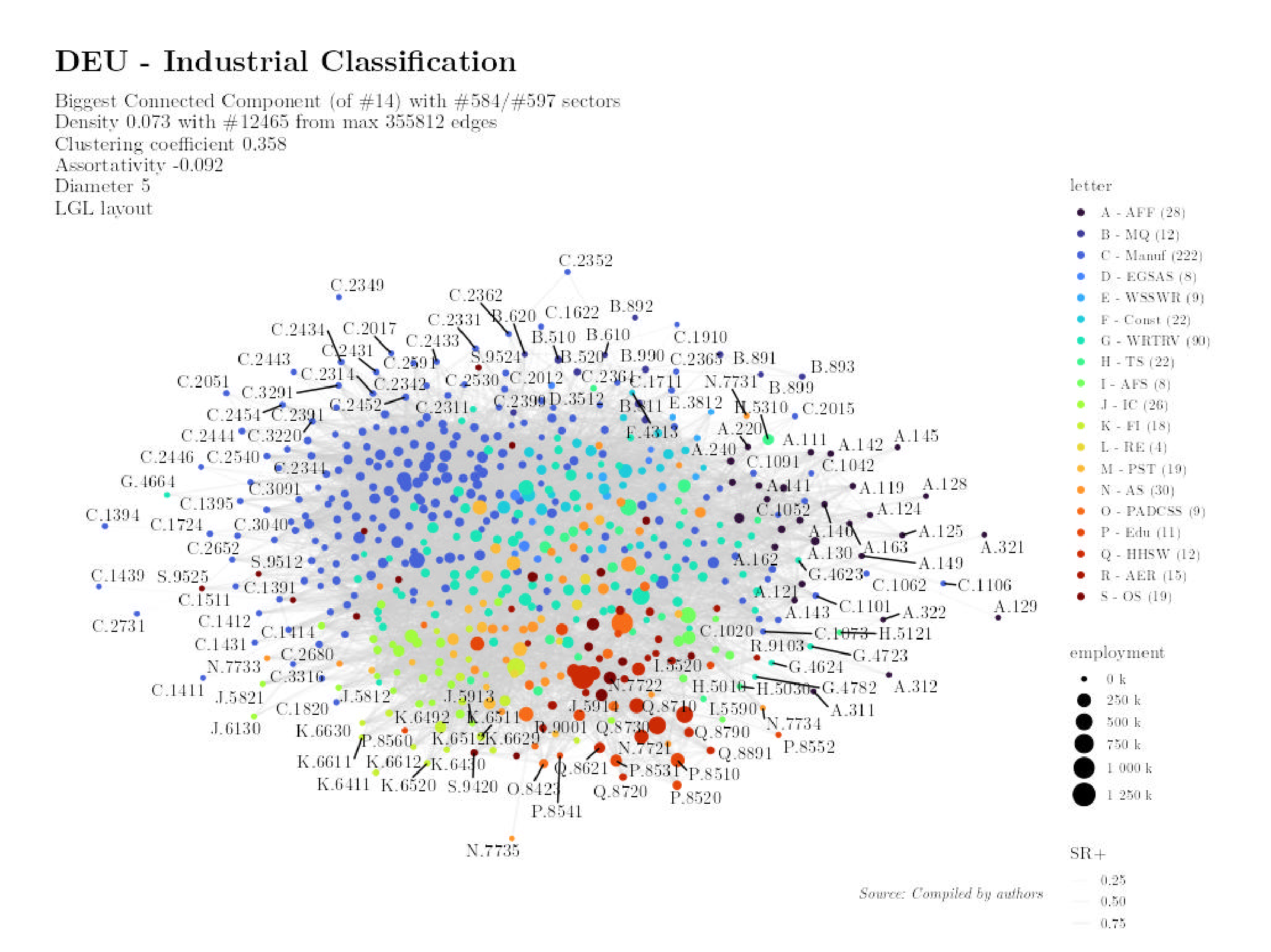}
    \label{fig:DEU-srn}
\end{figure}

\begin{figure}[ht]
    \caption{\footnotesize{Sweden Skill Relatedness Network 2007-2017. Colors represent ISIC rev. 4 equivalent economic industrial classification, numbers in color legend are number of sectors under the correspondent letter division. Node size represents average sectoral employment. Edge width represents $SR>0$.}}
    \centering
    \includegraphics[width=\pdfpagewidth]{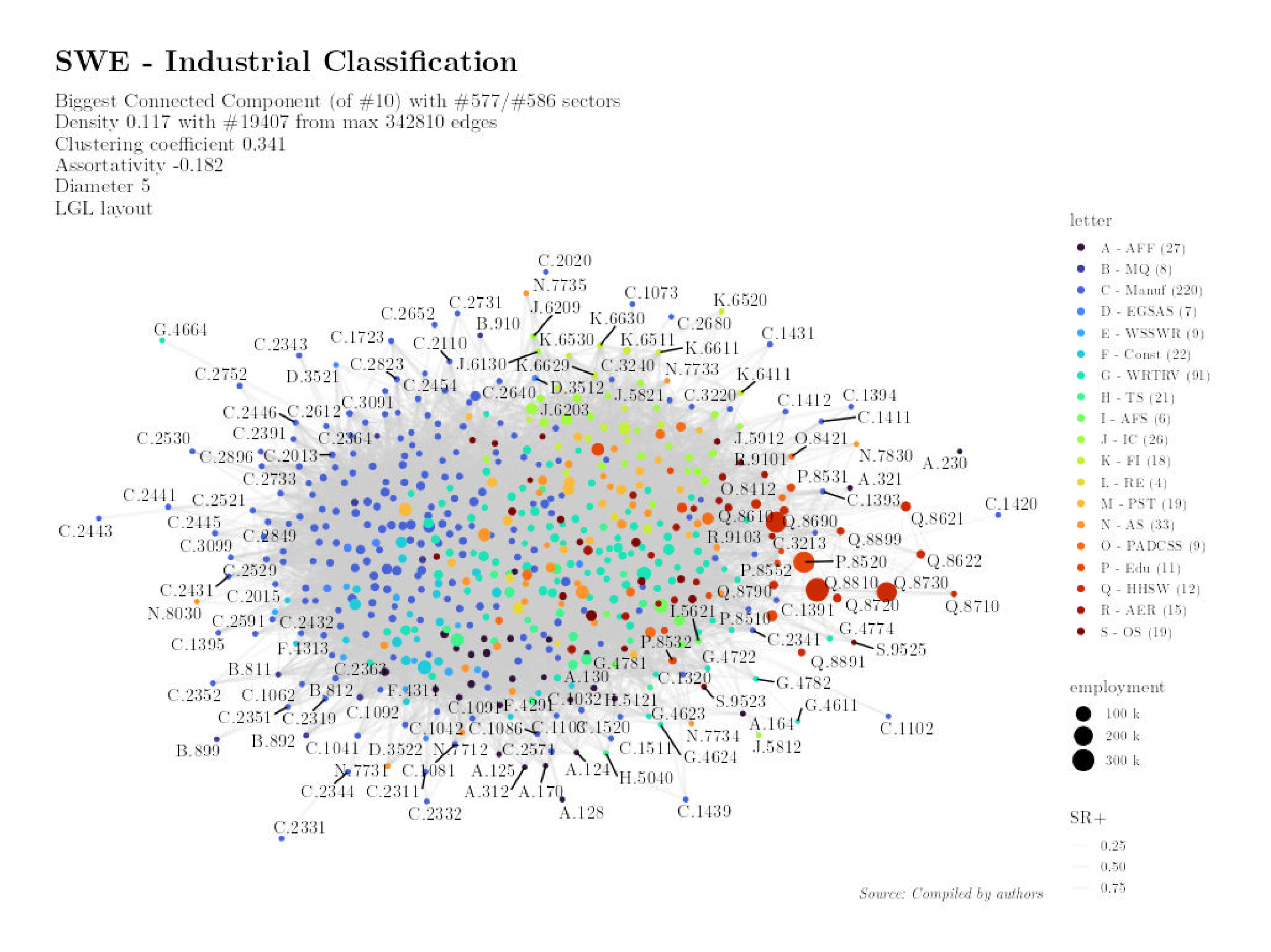}
    \label{fig:SWE-srn}
\end{figure}
\end{landscape}

\end{document}